\documentclass[aps,pra,twocolumn,superscriptaddress, nofootinbib]{revtex4-2}
\pdfoutput=1
\usepackage[utf8]{inputenc}
\usepackage[english]{babel}
\usepackage[T1]{fontenc}
\usepackage{amsmath}
\usepackage{hyperref}
\usepackage{tikz}
\usepackage{lipsum}
\usepackage{float}
\usepackage{graphicx}
\usepackage{latexsym}
\usepackage{amssymb}
\usepackage{nicefrac}
\usepackage{amsfonts}
\usepackage{enumerate}
\usepackage{upgreek}
\usepackage{subfigure}
\usepackage{bm}
\usepackage[version=3]{mhchem}
\usepackage{nicefrac}
\usepackage{bbold}
\usepackage{verbatim}
\usepackage{multirow}
\usepackage{color}
\usepackage{comment}
\usepackage{xcolor}
\usepackage{soul}

\usepackage{braket}

\usepackage{tikz}

\begin{document}
\title{Quantum Annealing with chaotic driver Hamiltonians}
\author{Henning Schl\"omer}
\affiliation{Department of Physics and Arnold Sommerfeld Center for Theoretical Physics (ASC), Ludwig-Maximilians-Universit\"at M\"unchen, Theresienstr. 37, M\"unchen D-80333, Germany}
\affiliation{Munich Center for Quantum Science and Technology (MCQST), Schellingstr. 4, D-80799 M\"unchen, Germany}
\affiliation{Department of Physics, Harvard University, Cambridge MA 02138, USA}
\author{Subir Sachdev}
\affiliation{Department of Physics, Harvard University, Cambridge MA 02138, USA}

\date{\today}
\begin{abstract}
Quantum annealing is a computational approach designed to leverage quantum fluctuations for solving large-scale classical optimization problems. Although incorporating standard transverse field (TF) terms in the annealing process can help navigate sharp minima, the potential for achieving a scalable quantum advantage for general optimization problems remains uncertain. Here, we examine the effectiveness of including chaotic quantum driver Hamiltonians in the annealing dynamics. Specifically, we investigate driver Hamiltonians based on a bosonic spin version of the Sachdev-Ye-Kitaev (SYK) model, which features a high degree of non-locality and non-commutativity. Focusing on MaxCut instances on regular graphs, we find that a considerable proportion of SYK model instances demonstrate significant speedups, especially for challenging graph configurations. Additionally, our analysis of time-to-solution scalings for the low autocorrelation binary sequence (LABS) problem suggests that SYK-type fluctuations can outperform traditional transverse field annealing schedules in large-scale optimization tasks.
\end{abstract}
\maketitle

\section{Introduction}
Physically-inspired annealing approaches aim to use fluctuations as a resource to solve classical optimization problems with highly complex energy landscapes and many local minima.  Historically, the concept of ``simulated annealing'' was proposed as a numerical tool to optimize a classical cost function with many independent parameters by performing Metropolis Monte Carlo simulations of an Ising model (whose energy landscape maps to the classical cost function~\cite{Albash2018}) at finite temperature. Systematically lowering the temperature by an annealing schedule then utilizes thermal fluctuations to escape local minima, ideally freezing the system in the ground state~\cite{Kirkpatrick1983}. 

Kadowaki and Nishimori proposed to replace thermal by quantum fluctuations~\cite{Kadowaki1998}, where a time-dependent transverse field (TF), called the driver Hamiltonian, is added to the Ising Hamiltonian that describes the optimization problem. Indeed, it has been established that quantum annealing shows convergence to the optimal (ground) state with larger probability than simulated annealing in a variety of cases if the same annealing schedule is used~\cite{Kadowaki1998, Brooke1999, farhi2000quantum, Marto2002, kadowaki2002study, farhi2002quantum, Santoro2002, Santoro_2006, Das2008}. The intuition for the enhanced performance is that quantum fluctuations allow for tunneling events through particularly spiky peaks of the energy landscape~\cite{Denchev2016}, which in contrast are not possible when using classical simulated annealing. 

The typical time-dependent Hamiltonian that evolves according to a specified annealing schedule $s(t)$ as a function of time $t$ reads
\begin{equation}
    \hat{H}_{\text{QA}} = s \hat{H}_{\text{C}} + (1-s) \hat{H}_{\text{D}}. 
    \label{eq:HQA}
\end{equation}
Here, $\hat{H}_{\text{C}}$ is the classical Ising Hamiltonian whose ground state corresponds to the solution of an optimization problem, and $\hat{H}_{\text{D}}$ is the driver Hamiltonian that, importantly, does not commute with $\hat{H}_{\text{C}}$. The annealing schedule $s(t)$ for a given total annealing time $T$ is chosen such that $s(0) = 0$ and $s(T) = 1$. Ideally, the time evolution is adiabatic, i.e. when starting in the ground state of the driver Hamiltonian, the system stays in the instantaneous ground state of the system and finally evolves into a state that has large overlap with the ground state of the target Hamiltonian. 

Naturally, the performance of the quantum annealing algorithm depends on the minimum (instantaneous) gap of the system during the evolution. In particular, according to the quantum adiabatic theorem~\cite{Born1928}, the annealing time $T$ relates to the minimum instantaneous energy gap $\Delta$ as $T \sim 1/\Delta^2$. Fully solving the time-dependent Schr\"odinger equation for Sherrington-Kirkpatrick (SK) spin glasses and evaluating their performance for different annealing times reveals strong correlations between instances of the SK model that are particularly hard to solve (from now on called ``hard instances'') and their corresponding minimum instantaneous energy gap~\cite{Rakcheev2023}. 

Following the definition in~\cite{Zhou2020}, such a hard instance is characterized by a ``diabatic bump'', where the success probability to end up in the ground state of the optimization Hamiltonian features a local maximum at intermediate annealing times, before increasing towards unity in the adiabatic regime $T\sim 1/\Delta^2$ (see also Ref.~\cite{Rakcheev2023}). Intuitively, diabatic transitions that happen at short annealing times partially deplete the ground state occupancy \textit{before} the instantaneous gap minimum, hence reducing the transition probability from the ground to the excited state at the gap closing. In contrast, for longer annealing times (but shorter than needed to be in the adiabatic regime), the ground state occupancy before the instantaneous gap minimum stays constant, and transitions to the first excited states are highly probable when approaching the minimum energy gap.

Recent technological advances in designing and engineering special purpose quantum annealing machines has sparked additional interest in quantum annealing, with the ultimate goal of building a scalable quantum device that implements coherent and adiabatic time evolution for Hamiltonians with flexible interactions~\cite{Hauke2020}. With major efforts invested in superconducting qubit devices~\cite{Johnson2011, Dickson2013, Denchev2016, Weber2017, King2022, Willsch2022, King2023}, neutral atom~\cite{Glaetzle2017, Ebadi2022} and trapped ion~\cite{Islam2013, Richerme2013} annealers are promising alternative architectures offering several advantages, including long coherence times and all-to-all connectivities. 
 
The transverse field driver $\hat{H}_{X} = \sum_i \hat{\sigma}^x_i$ (with $\hat{\sigma}^x_i$ the Pauli-$x$ matrix on site $i$) constitutes the Hermitian complete combination of operators that couple all binary sequences of the classical problem, and has established as the paradigmatic choice both in theoretical and experimental considerations~\cite{Hauke2020}. However, a transverse field only couples states with Hamming distance $d_H = 1$, i.e., application of the driver results in a superposition of single spin flips for a given product state. This, in turn, opens the question whether driver Hamiltonians that include (non-local) multi-point interactions of degree $>1$ may help to escape local minima and anneal to the ground state more efficiently in complicated optimization landscapes. 

Moreover, the transverse field is a ``stoquastic'' Hamiltonian, which results in all path integral configurations to have real and positive weight, ultimately rendering the system sign-problem free~\cite{landau2021guide}. Hence, properties of the annealing Hamiltonian can be efficiently calculated using classical techniques. Though Monte Carlo is designed to simulate the equilibrium Boltzmann distribution (and hence only equilibrium properties can be directly computed), it has been shown that it can capture features of quantum dynamics and hence of the annealing process~\cite{Isakov2016, Denchev2016, Jiang2017}. Indeed, identical scalings of computational time with system size for genuine and Monte Carlo simulated quantum annealing have been reported~\cite{Isakov2016, Denchev2016, Jiang2017}. This may suggest that if an instance of a problem is particularly hard to solve using classical Monte Carlo methods, it is also hard to solve using quantum annealing with a TF driver. Utilizing driver Hamiltonians that can not be efficiently simulated classically may hence enable performance enhancements in particular for hard instances~\cite{Albash2019}.

The effects of using non-stoquastic annealing schedules have been numerically analyzed in Refs.~\cite{crosson2014different, Hormozi2017}. Hormozi \textit{et al.} demonstrated that adding non-stoquastic terms to the driver Hamiltonian can enhance the performance of hard instances by promoting diabatic transitions at short annealing times (i.e. away from the adiabatic regime)~\cite{Hormozi2017}. Crosson \textit{et al.} reported similar findings, demonstrating that non-stoquastic terms do enhance the success rate, although not significantly more than the addition of stoquastic terms~\cite{crosson2014different}.
Furthermore, it was shown that by including non-stoquastic drivers that add antiferromagnetic quantum fluctuations to the annealing schedule, first order phase transitions can be reduced to be of second order for specific problems~\cite{Seki2012, Seoane2012, Seki2015, Nishimori2017}, enhancing the time to solution from scaling exponentially to polynomially with system size. From a different perspective, it has been demonstrated that tailored (generally non-stoquastic) Hamiltonians can be used to solve optimization problems with additional constraints, whereby the driver Hamiltonian is to be constructed such that it commutes with the constraint operators~\cite{Hen2016_1, Hen2016}. 

While progress has been made in understanding the effect of certain additional terms in the driver Hamiltonian, analyzing which type of fluctuations leads to the best performance for general optimization problems remains an open question. Following intuitive reasoning one may argue that the inclusion of non-local terms that couple states with larger Hamming distance $d_H>1$ can allow for a more efficient exploration of the optimization landscape, hence leading to a computational advantage compared to the regular transverse field. This is in particular motivated by recent progress in neutral atom and trapped ion setups, which allow for the implementation of long-range interactions~\cite{Bluvstein2022, bruzewicz2019trapped}. Additional quantum fluctuations by making the driver Hamiltonian itself quantum may further enhance the annealing performance. This motivates us to study optimization performances when implementing Sachdev-Ye-Kitaev-type fluctuations into the annealing schedule, whose nature are highly chaotic. We demonstrate that drastic computational enhancements can be achieved for general hard optimization problems.

We will introduce the bosonic SYK model and its annealing schedules in Section~\ref{sec:bSYK}. The application to the MaxCut optimization is presented in Section~\ref{sec:maxcut}, and to the LABS problem in Section~\ref{sec:labs}. Finally Appendix~\ref{sec:App1} describes a simple solvable toy model which illustrates the advantage of non-locality in quantum annealing.

\section{The (bosonic) SYK model}
\label{sec:bSYK}
The Sachdev-Ye-Kitaev (SYK) model~\cite{YeSachdev1993, Kitaev_lecture} is a strongly correlated, chaotic quantum many-body system that features highly entangled low-energy states, and has provided broad insights ranging from non-Fermi liquids in solid state materials~\cite{Chowdhury2022} to charged black holes with 2D anti-de Sitter horizons~\cite{Sachdev2010, Sachdev2015}. Specifically, the SYK model constitutes a concrete model Hamiltonian that features many-body quantum chaos, which is characterized by fast scrambling of quantum information~\cite{Yasuhiro2008, Shenker2014, Hosur2016} that saturates universal bounds (formalized by the Lyapunov exponent)~\cite{Maldacena2016, Kobrin2021}. 

The particle-hole symmetric Majorana formulation of the SYK model with $q$-local interactions (SYK$_{q}$) reads
\begin{equation}
    \hat{H}_{\text{SYK}_{q}} = i^{q/2} \sum_{i_1 < \dots < i_q} J_{i_1 \dots i_q} \hat{\chi}_{i_1} \dots \hat{\chi}_{i_q}. 
\end{equation}
Here, $\hat{\chi}_i$ are $N$ (hermitian) Majorana operators and $J_{i_1 \dots i_q}$ are independent Gaussian random variables with zero mean and variance 
\begin{equation}
    \text{Var}[J_{i_1\dots i_q}] = \frac{(q-1)!}{N^{q-1}}.
\end{equation}
The latter ensures that the bandwidth of the model is of order unity for all $N$. 

There has been increased recent interest in studying bosonic versions\footnote{By bosonic we specifically mean that the system's Hilbert space has a tensor product structure whereby operators from different sites commute.} of the SYK model, in particular due to their potential to be simulated with quantum computers~\cite{Tezuka2024, Swingle2024}. Following Refs.~\cite{Berkooz2018, Berkooz2019, Swingle2020}, we define the bosonic SYK model with $q$-local all-to-all interactions (from now on denoted by bSYK$_{q}$)\footnote{We note that a variety of bosonic versions of the SYK model have been proposed, including a two-component spin model~\cite{Tezuka2024}. However, we expect that our results largely don't depend on the details of the spin model; see also the discussion in Ref.~\cite{Swingle2024}},
\begin{equation}
    \hat{H}_{\text{bSYK}_{q}} = \sum_{i_1 < \dots < i_q}\sum_{\alpha_1, \dots, \alpha_q} J_{i_1 \dots i_q}^{\alpha_1 \dots \alpha_q} \hat{\sigma}_{i_1}^{\alpha_1} \dots \hat{\sigma}_{i_q}^{\alpha_q}. 
    \label{eq:bSYK}
\end{equation}
Here, $\alpha_i = x,y,z$; $J_{i_1 \dots i_q}^{\alpha_1 \dots \alpha_q}$ are again (real) random Gaussian variables with vanishing mean and variance
\begin{equation}
    \text{Var}[J_{i_1\dots i_q}^{\alpha_1 \dots \alpha_q}] = \frac{(q-1)!}{N^{q-1}}.
\end{equation}
It has been shown that many of the defining features of the SYK model, such as power law correlation functions and an extensive low-temperature entropy, are present also in the above bosonic formulation~\cite{Swingle2024}. Furthermore, at large $q$, the bSYK$_q$ model features the same four-point functions as those of the SYK$_q$ Hamiltonian, leading to maximal chaos at low temperature~\cite{Swingle2024}. This makes it an attractive candidate to study  emergent gravitational descriptions and quantum chaos using quantum computers, primarily because it removes the issue of non-local operators when mapping fermionic degrees of freedom to spins.

With implementations of (variants of) the SYK model becoming more realistic, we here ask a different question: Can the chaotic and fast scrambling nature of the system enhance quantum annealing approaches to solve classical optimization problems? Building on the intuition above, non-locality combined with a high degree of non-commutativity of the SYK model may lead to higher probabilities to escape local minima and find the global minimum of the energy landscape. In Appendix~\ref{sec:App1}, we present a simple two-qubit energy landscape that illustrates the advantage of the above on an intuitive basis.  

In the following, we will focus on two ways of including SYK-type driver Hamiltonians into the annealing schedule:

\begin{enumerate}[(S1)]
    \item  First, we study an annealing process where we start in the ground state of the bosonic SYK model, which is then annealed to the classical optimization Hamiltonian, i.e., 
    \begin{equation}
        \tag{S1}
        \hat{H}_{\text{QA}}^{(1)} = s \hat{H}_{\text{C}} + (1-s)\hat{H}_{\text{bSYK}_q}, 
        \label{eq:HQA_1}
    \end{equation}
    with linear ramp $s(t) = t/T$. We note that the ground state preparation of $\hat{H}_{\text{bSYK}_q}$ is experimentally challenging. However, including the chaotic driver in various ways allows us to compare different settings of adding non-local fluctuations to the annealing process.

\item Second, we study the performance when adding SYK-type fluctuations on top of the TF schedule, 
    \begin{equation}
    \tag{S2}
    \label{eq:HQA_2}
    \begin{aligned}
        \hat{H}_{\text{QA}}^{(2)} = s \hat{H}_{\text{C}} + &(1-s) \hat{H}_{X} \\&+ 4s(1-s)\hat{H}_{\text{bSYK}_q}. 
    \end{aligned}
    \end{equation}
    In particular, here the system starts in the product state of the transverse field at $s=0$. Fluctuations of the SYK-type model are then added to the Hamiltonian (where the corresponding strengths of the TF and SYK model are equal at $s=1/2$), ending at the classical Ising Hamiltonian at $s=1$ (see also Refs.~\cite{crosson2014different, Hormozi2017}). This approach is specifically motivated by the design of hardware annealers, where the system is initialized in a product state and then evolves dynamically under a time-dependent Hamiltonian, which can potentially be engineered to include more terms beyond the transverse field.
\end{enumerate}

\begin{figure*}
\centering
\includegraphics[width=0.87\textwidth]{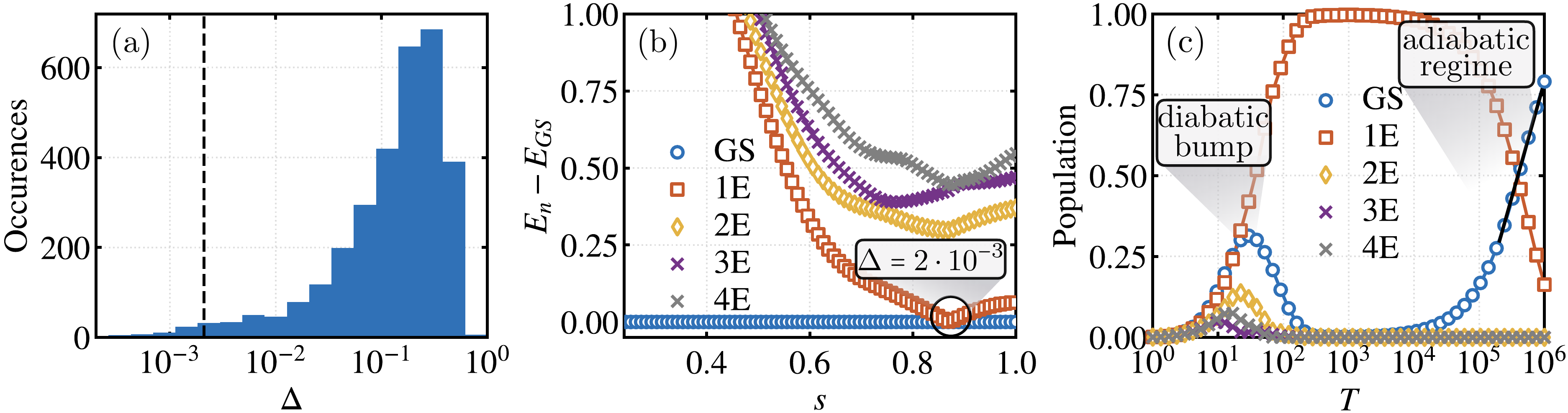}
\caption{\textbf{MaxCut with transverse field driver.} (a) Distribution of minimum instantaneous energy gaps $\Delta$ during adiabatic evolution under annealing schedule~\eqref{eq:HQAX} for $N_g = 3050$ random instances of 3-regular MaxCut graphs with $N=14$ vertices. (b) The instantaneous eigenspectrum (relative to the ground state energy) as a function of annealing parameter $s$ for a single hard graph instance. The corresponding minimum instantaneous energy gap of $\Delta \approx 2 \cdot 10^{-3}$ at $s\approx 0.85$ is indicated by the black dashed line in the distribution of $\Delta$ in (a). (c) Decomposition of the dynamically evolved state with annealing time $T$ into the eigenstates of the MaxCut Ising Hamiltonian. Hard instances are characterized by a diabatic bump of the ground state weight at small times. At times $T \gtrsim \Delta^{-2}$, the ground state weight approaches unity exponentially (adiabatic regime, indicated by the black solid line).}
\label{fig:oX}
\end{figure*}

Both schedules~\eqref{eq:HQA_1} and~\eqref{eq:HQA_2} are compared to the standard TF annealing approach,
\begin{equation}
\tag{TF}
    \hat{H}_{\text{QA}}^X = s\hat{H}_C + (1-s) \hat{H}_X.
    \label{eq:HQAX}
\end{equation} 
Specifically, we test the performance of SYK-type drivers for two classical optimization problems: In Section~\ref{sec:maxcut}, we look at MaxCut problems on random $d$-regular graphs, which can be mapped to an Ising model with 2-body interactions on the respective graphs. We demonstrate that for hard optimization problems, the SYK driver has a finite probability of performing significantly better than the TF driver. In Section~\ref{sec:labs}, we look at the Low Autocorrelation Binary Sequence (LABS) problem, which aims to find a sequence of binary numbers with minimal autocorrelations. In this case, the classical optimization problem can be encoded in an Ising Hamiltonian that consists of non-local 4-body terms. All classical (exact and heuristic) algorithms known scale exponentially with system size for LABS. By solving the time-dependent Schr\"odinger equation for small system sizes, our results suggest that using chaotic drivers may lead to a scaling advantage compared to conventional annealing schedules.

\section{MaxCut}
\label{sec:maxcut}

Our first test case is the MaxCut optimization problem, where one aims to maximize the number of weighted edges (given by the set $E=\{(\braket{i,j}, w_{ij})\}$) in a graph that are ``cut'' by a given partition of the vertices ($V = {1,2,\dots, N}$) into two sets. For a given graph with edges $E$ and vertices $V$, the problem can be mapped to finding the ground state $\hat{H}_{\text{MC}} \ket{\Psi_0}= E_{\text{GS}}\ket{\Psi_0}$ of the following Ising Hamiltonian,
\begin{equation}
    \hat{H}_{\text{MC}} = \sum_{\braket{i,j}} w_{ij} \hat{\sigma}_i^z \hat{\sigma}_{j}^{z}.
    \label{eq:MaxCut}
\end{equation}
Here, $\hat{\sigma}_i^z$ is the $z$-component of the spin-1/2 Pauli operator. While finding exact solutions for general graphs is exponentially hard with increasing system size $N$, many heuristic algorithms have been proposed. 
Here, one aims to design an algorithm that finds a bit string $\mathbf{z}$ with energy $E(\mathbf{z})$ such that 
\begin{equation}
    E(\mathbf{z})/E_{\text{GS}} \geq r^*.
\end{equation}
Indeed, it has been rigorously shown that beyond a certain approximation ratio $r^*$, MaxCut is $NP$-hard~\cite{Hastad2001}. 
The currently best known algorithm is the one by Goemans and Williamson, with $r^* \approx 0.88$~\cite{Goemans1995}. When restricting the edges of the graph to be bimodal, this bound is raised to $r^* \approx 0.93$~\cite{Halperin2004}. Farhi, Goldstone, and Gutman showed that the quantum approximate algorithm (QAOA) has a guaranteed minimum approximation ratio of $r^* \gtrsim 0.69$, and can in principle reach $r^* = 1$ for infinite depth~\cite{farhi2014quantum}. Similarly, $r^* \rightarrow 1$ for the quantum annealing approach at infinite annealing times, which makes MaxCut a promising testing ground for a potential quantum advantage~\cite{farhi2019quantum}.

Here, we focus on the performance of quantum annealing when using both the paradigmatic transverse field driver Hamiltonian as well as bSYK$_{q}$ drivers with schedules~\eqref{eq:HQA_1} and \eqref{eq:HQA_2}. We shall restrict our considerations to hard instances of the MaxCut problem on $d$-regular graphs, where every vertex is connected to $d$ other vertices; specifically, we choose $d=3$.

\begin{figure*}
\centering
\includegraphics[width=0.85\textwidth]{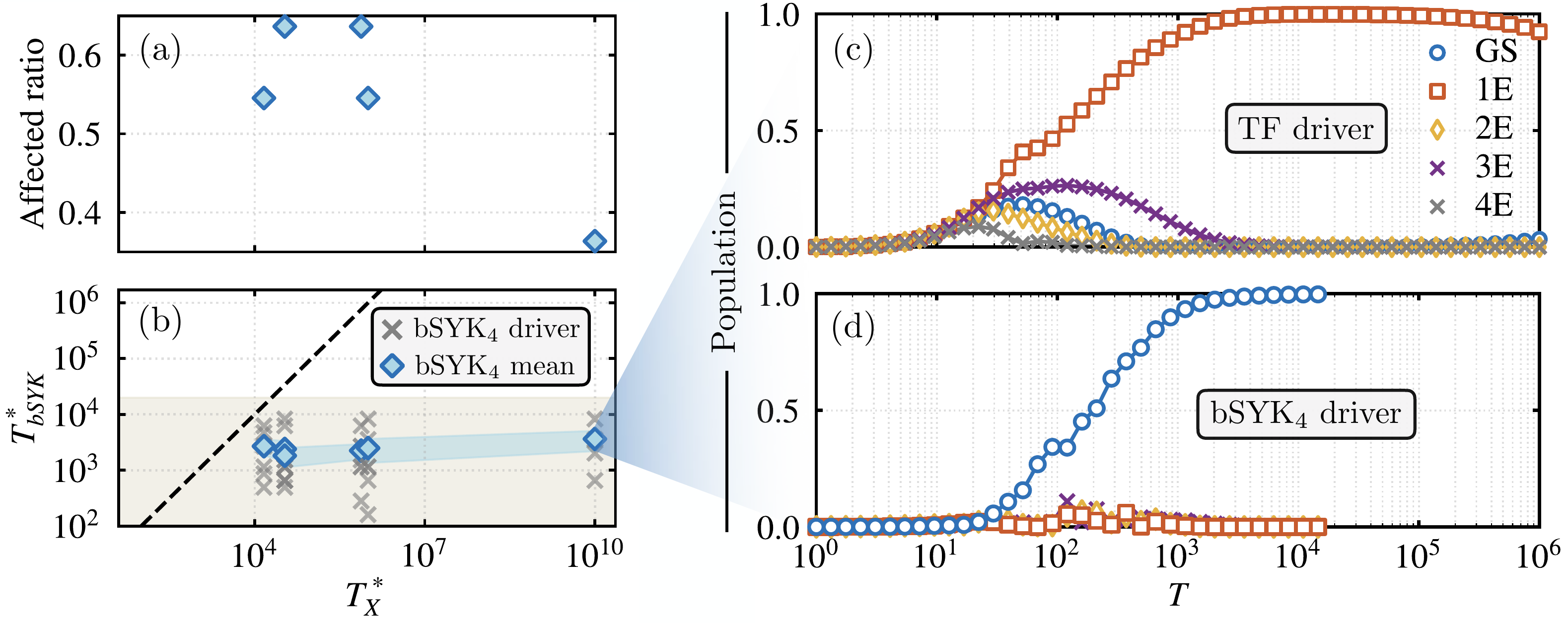}
\caption{\textbf{TF vs. bSYK drivers with annealing schedule \eqref{eq:HQA_1}.} (a) The ratio of bSYK$_4$ instances that lie within the considered numerical annealing time window and that show better performance compared to the TF driver. (b) $T^*_{\text{bSYK}_{4}}$ for the successful instances (grey dots) in comparison to $T^*_X$ when using the TF driver. Mean values of $T^*$ for the bSYK$_4$ driver are shown with blue diamonds; the shaded blue region illustrates the error to the mean. The dashed line is a guide to the eye signaling $T^*_X = T^*_{\text{bSYK}_{4}}$, and the beige region corresponds to the range we consider numerically, i.e., $T^*_{\text{bSYK}_{4}}\leq 2\cdot 10^4$. (c) Annealing performance for a particularly hard instance using the TF model. Around the longest considered annealing times, the ground state population is barely entering the adiabatic regime. (d) An exemplary successful bSYK$_4$ instance showing significant enhancement of the annealing performance for the same MaxCut problem as considered in (c). As parity is not conserved for the bSYK driver, the shown ground state population in (d) corresponds to the sum over both minimal energy parities of the classical MaxCut Hamiltonian; in (c), only one parity sector is shown.}
\label{fig:comp}
\end{figure*}

For a given graph size $N=14$, we draw $N_g = 3050$ random instances, where the weights $w_{ij}$ are randomly chosen from a uniform distribution $\in [0,1]$. To isolate the hard instances, we calculate the instantaneous eigenspectrum of the annealing schedule using the TF driver~\eqref{eq:HQAX} as a function of annealing parameter $s$ and extract the minimum instantaneous energy gap $\Delta$. The resulting distribution of $\Delta$ is shown in Fig.~\ref{fig:oX}~(a). 
While most graph instances have an instantaneous gap $\Delta >10^{-2}$, we isolate those configurations with $\Delta < 10^{-2}$. As an example, we show the instantaneous eigenspectrum decomposition for one random graph in Fig.~\ref{fig:oX}~(b). The minimum instantaneous gap is $\Delta \approx 2 \cdot 10^{-3}$, which appears at $s\approx 0.85$. The corresponding value of $\Delta$ of the shown graph is underlined by the black dashed line in the distribution in Fig.~\ref{fig:oX}~(a).

For all graph instances with small instantaneous gaps $\Delta < 10^{-2}$, we calculate the annealing performance for varying annealing time $T$. We note that the Hamiltonian~\eqref{eq:HQAX} commutes with the parity operator $\hat{P} = \sum_{i}\hat{\sigma}^x$, which we conserve during our simulations. For the random seed presented in Fig.~\ref{fig:oX}~(b), annealing results are shown for $10\leq T\leq 10^{6}$ in Fig.~\ref{fig:oX}~(c). 

A typical hard instance is characterized by a diabatic bump for short annealing times $T$, where diabatic transitions during the annealing sweep lead to enhanced ground state populations by the end of the protocol. For times $T>\Delta^{-2}$, the annealing schedule reaches the adiabatic regime, whereby the ground state population exponentially approaches unity [indicated by the black solid line in Fig.~\ref{fig:oX}~(c)]. These observations are in line with previous definitions of hard instances of MaxCut problems, see e.g.~\cite{Zhou2020, Rakcheev2023}. 

For hard instances, we now compare the annealing performance of the TF driver to schedules~\eqref{eq:HQA_1} and~\eqref{eq:HQA_2} that include the bosonic SYK model in the driver Hamiltonian. In contrast to the TF schedule~\eqref{eq:HQAX}, we need to solve the time-dependent Schr\"odinger equation for varying annealing times $T$ in the full Hilbert space of $N$ spins, as no symmetries can be exploited for the SYK model; furthermore, the matrices are dense, such that methods for sparse Hamiltonians can't be used. Therefore, in the following we focus on six representative hard graph instances, and analyze the annealing performance for $N_{\text{bSYK}_q} = 12$ SYK realizations for annealing times $10\leq T \leq 2 \cdot 10^4$. In particular, we use Runge-Kutta and exact diagonalization methods available in the \texttt{QuSpin} package~\cite{Quspin1, Quspin2}.

\begin{figure*}
\centering
\includegraphics[width=0.87\textwidth]{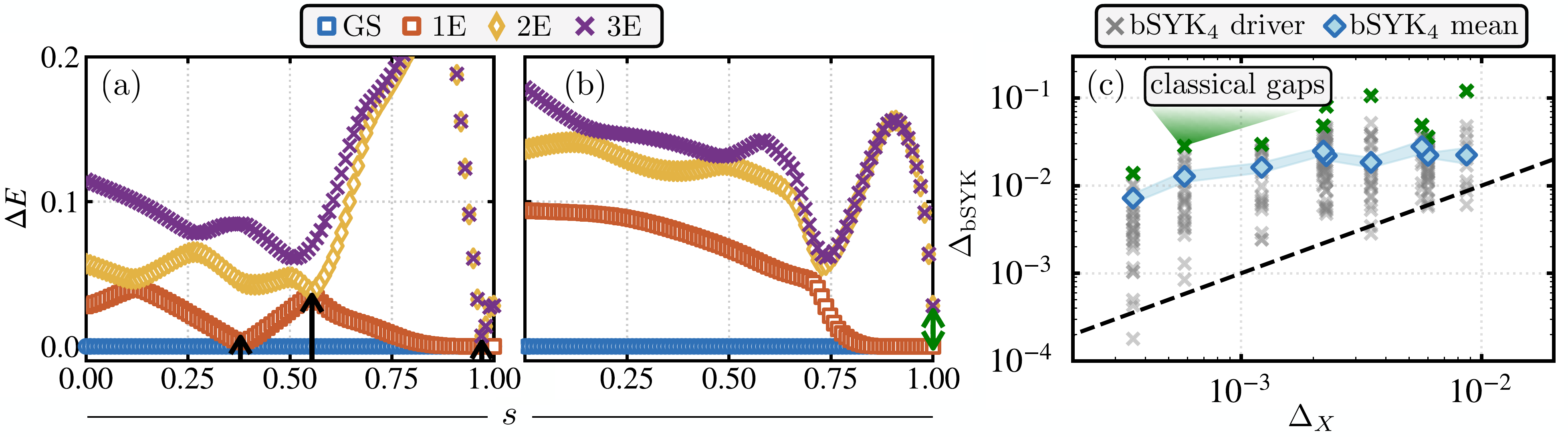}
\caption{\textbf{Instantaneous spectra and minimum gaps for annealing schedule~\eqref{eq:HQA_1}.} (a) and (b) show the instantaneous eigenspectrum of~\eqref{eq:HQA_1} as a function of annealing parameter $s$ for two random bSYK$_4$ instances. In (a), a series of avoided crossings (indicated by the left and center arrow) as well as a small instantaneous gap $\Delta_{\text{bSYK}_q} = E_2 - E_{\text{GS}}$ (right arrow) can lead to a depletion of the ground state population. (b) shows the same random bSYK$_4$ instance as in Fig.~\ref{fig:comp}~(d), where a wide gap opening of the instantaneous spectrum leads to an ideal annealing performance. The classical gap of the MaxCut optimization problem at $s=1$ is indicated by the green arrow. (c) Comparison of the minimum instantaneous energy gaps when using the TF and bSYK$_4$ driver. Grey crosses show individual instances of the SYK-type driver, solid diamonds show corresponding mean values. Green crosses indicate the classical gaps at $s=1$, which is an upper bound for the instantaneous gap $\Delta_{\text{bSYK}_4}$. Due to different conserved symmetries, $\Delta_{\text{bSYK}_4} = E_2 - E_{\text{GS}}$, while $\Delta_{X} = E_1 - E_{\text{GS}}$ (see main text). In almost all cases, instantaneous gaps are significantly opened.}
\label{fig:comp_gap}
\end{figure*}

As a metric to compare the annealing performance of the TF and bSYK driver, we extract the time $T$ for which the ground state population reaches a value $p_{\text{GS}}(T^*) = 0.9$ after the diabatic bump. We note that this is in contrast to e.g. Ref.~\cite{Hormozi2017}, where a short fixed annealing time of $T = 100$ was chosen for all graph instances. In particular, our metric relies entirely on adiabadicity of the annealing schedule, i.e., enhanced success probabilities due to promoted diabatic transitions before the gap closing are excluded. 

\underline{\textit{Annealing schedule \eqref{eq:HQA_1}}.} We start by analyzing the case where the initial state of the annealing schedule is the ground state of the bSYK$_q$ model, i.e., the Hamiltonian as a function of annealing parameter $s$ is $\hat{H}_{\text{QA}}^{(1)} = s \hat{H}_{\text{MC}} + (1-s)\hat{H}_{\text{bSYK}_q}$. We denote the bSYK$_4$ realizations which result in a  $T^{*}_{\text{bSYK}_4}$ that is lower than $T^*_X$ and below our numerical time evolution threshold of $2\cdot 10^4$ as successful instances.  For $q=4$, the fraction of successful instances is shown in Fig.~\ref{fig:comp}~(a) as a function of $T^*_X$ for various hard graph instances. 
For graphs with solution times $T^*_X \lesssim 10^6$, success ratios are relatively constant and range between $\sim 55-65\%$. For the hardest considered instance, the success ratio is found to be considerably lower, with $\sim 35\%$ of the bSYK realizations leading to faster annealing. 

For the successful instances, times $T^*_{\text{bSYK}_4}$ are compared to corresponding times $T^*_X$ when using the TF driver in Fig.~\ref{fig:comp}~(b). Grey crosses show results for various bSYK$_4$ driver realizations, blue dots show the mean over the successful instances. The black dashed line shows where $T^*_{\text{bSYK}_4} = T^*_X$, and the beige shading displays our numerically considered range of $T^*_{\text{bSYK}_4}\leq 2 \cdot 10^4$. Indeed, successful instances are seen to result in solution times that stay approximately constant of the order $\bar{T}^*_{\text{bSYK}_4} \sim \mathcal{O}(10^3)$, independent on the solution time $T^*_{\text{X}}$ when using the TF driver.

This is corroborated in Fig.~\ref{fig:comp}~(c), which shows the eigenstate populations of the MaxCut problem for varying $T$ for a particularly hard instance using the TF driver. Around the largest considered annealing times, the TF driver barely starts to gain a visible ground state population. In stark contrast, for the bSYK$_4$ driver, the optimal solution is found with high probability $p(T^*) = 0.9$ already for $T^*_{\text{bSYK}_{4}} \approx 2 \cdot 10^3$ [Fig.~\ref{fig:comp}~(d)]. We crudely extrapolate $T^*_X$ to be $10^{10}$ for the shown example, which corresponds to the outer right data point in Fig.~\ref{fig:comp}~(b). The exact time $T^*_X$ is, however, only of secondary importance, the takeaway being that if an instance of the bSYK model is successful, it allows to reach large ground state probabilities on significantly shorter time scales. However, we note that smaller success ratios are observed for the hardest considered optimization problem (though the small sample size of SYK instances lead to only coarse resolutions). 

The behavior of $T^*_{\text{bSYK}_4}$ as a function of $T^*_X$ suggests that a critical level of graph difficulty exists where the bSYK driver starts to outperform the mere use of TF fluctuations. For the considered 3-regular MaxCut problem with $N=14$ vertices, we read off an approximate threshold of $T^*_X \sim 10^3$, see Fig.~\ref{fig:comp}~(b). This aligns with the intuition given above: For comparably easy instances with large gaps, the transverse field efficiently navigates through the optimization landscape and quickly finds optimal solutions. However, for hard instances with complex energy landscapes, the non-local nature of the SYK model offers an advantage over the TF, whereby it can find the global minimum more efficiently.

\begin{figure*}
\centering
\includegraphics[width=0.87\textwidth]{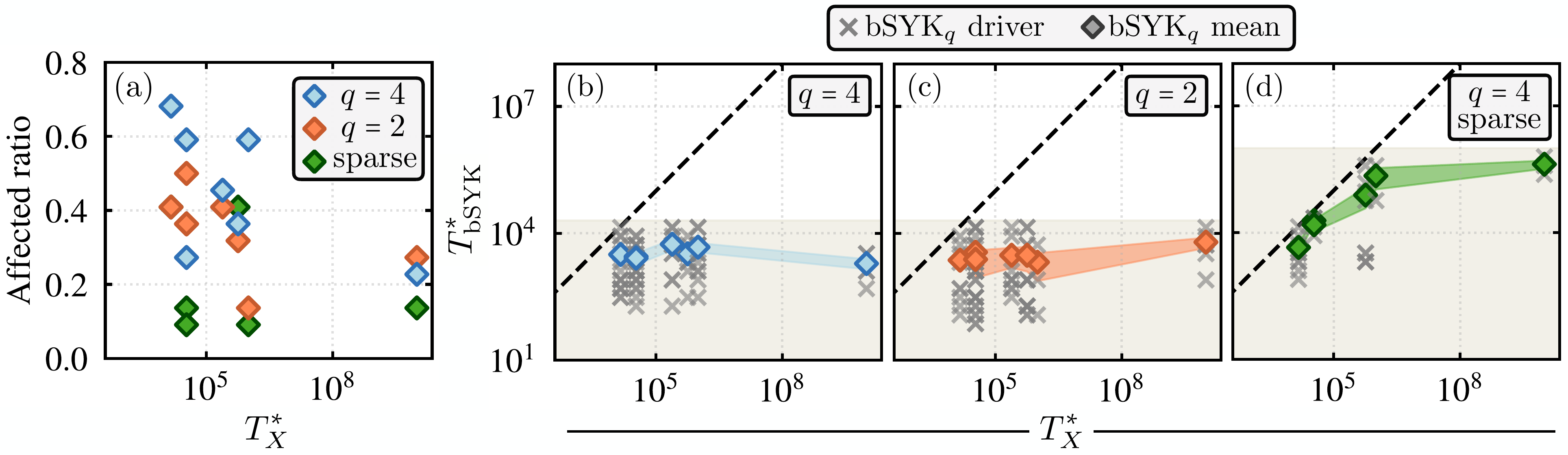}
\caption{\textbf{TF vs. bSYK drivers with annealing schedule \eqref{eq:HQA_2}.} (a) The ratio of instances that lie within the considered numerical annealing time window and that show better performance compared to the TF driver. We consider $N_{\text{bSYK}_q} = 24$ driver instances for each hard graph, for $q=4$ (blue), $q=2$ (orange), and a sparsified bSYK$_4$ model that keeps only $4N$ interactions terms (green). (b) $T^*$ for the successful instances (grey dots) in comparison to TF drivers for $q=4$. Mean values of $T^*$ for successful instances are shown in blue. The dashed line indicates where $T_{\text{bSYK}_q}^* = T_{X}^*$, and the beige shading shows our numerically considered annealing time range for the bSYK driver. The performance of driver~\eqref{eq:HQA_2} is comparable to~\eqref{eq:HQA_1}, whereby the average $T^*$ over successful instances remains approximately constant. (c) $T^*$ for $q=2$. A larger variation of $T^*$ leads to slightly lower success ratios, see (a). Nevertheless, the performance is comparable to the case where $q=4$. (c) $T^*$ for the sparse bSYK$_4$ model. Only very few driver realizations lead to an enhanced annealing performance, with notably larger $T^*$ for most MaxCut graphs.}
\label{fig:comp_4s}
\end{figure*}

The enhanced adiabatic performance when using the SYK-type driver suggests that the gap of the instantaneous eigenspectrum is widened (see also the simple toy model we present in Appendix~\ref{sec:App1}). Fig.~\ref{fig:comp_gap}~(a) and (b) show the instantaneous eigenspectrum of the annealing schedule~\eqref{eq:HQA_1} for two different random instances of the bSYK driver for the hardest considered MaxCut graph, cf. Fig.~\ref{fig:comp}. As the bosonic SYK Hamiltonian does not commute with $\hat{P} = \sum_i \hat{\sigma}_i^x$, it can couple states of different parities $\hat{P}$. 
This is in contrast to the TF driver, which preserves the parity during time evolution. Hence, in our numerical time evolution, the final states at $s=1$ are two-fold degenerate in the case of the bSYK driver, as seen in Fig.~\ref{fig:comp_gap}~(a) and (b). Correspondingly, by the end of the annealing schedule, the ground state population corresponds to the sum over the degenerate ground state subspace.

The ground state population can thus be depleted via two main mechanisms: (i) The gap between the ground and second excited state closes, as seen exemplary at the right arrow in Fig.~\ref{fig:comp_gap}~(a). (ii) The appearance of a series of avoided crossings between the ground, first and second excited state, see the left and center arrow in Fig.~\ref{fig:comp_gap}~(a). However, we find that in many driver realizations, neither of the above appear and instantaneous gaps are significantly opened. Fig.~\ref{fig:comp_gap}~(b) shows the spectrum for the bSYK driver realization presented in Fig.~\ref{fig:comp}~(d). Indeed, there is no avoided crossing of the ground and first excited state and a large instantaneous gap is present throughout the whole range of $0\leq s \leq 1$. The minimum instantaneous energy gap appears at $s=1$, i.e., it is determined by the classical gap of the MaxCut spectrum itself [see the green arrow in Fig.~\ref{fig:comp_gap}~(b)]. This, in turn, leads to an ideal annealing performance presented in Fig.~\ref{fig:comp}~(d). 

In Fig.~\ref{fig:comp_gap}~(c), we compare the minimum instantaneous energy gaps $\Delta_{\text{bSYK}_q} = \text{min}_s (E_2 -E_{\text{GS}})$ when using schedule~\eqref{eq:HQA_1} to $\Delta_{X} = \text{min}_s (E_1 -E_{\text{GS}})$ when using the TF driver~\eqref{eq:HQAX}, for nine hard instances and 50 bSYK realizations each. 
We see that for almost all instances of the SYK model, the minimum instantaneous energy gap is indeed significantly widened. Mean values are shown by blue diamonds, which, in line with calculations of $T^*$ in Fig.~\ref{fig:comp}~(b), stay almost constant independent of $\Delta_X$. Corresponding classical gaps of the MaxCut problems (i.e. the instantaneous gap at $s=1$) are shown by green crosses, which is an upper bound of $\Delta_{\text{bSYK}_q}$ that indeed is saturated for a finite number of bSYK$_4$ driver instances. Nevertheless, we again note that a series of avoided crossings can lead to a depletion of the ground state population, even for large minimum instantaneous gaps $\Delta_{\text{bSYK}_q}$. As a result, this leads to lower fractions of bSYK instances to be successful in our annealing simulations in Fig.~\ref{fig:comp}.

\underline{\textit{Annealing schedule \eqref{eq:HQA_2}}.} Experimental preparation of the bSYK$_q$ ground state with high fidelity is a very challenging task. Furthermore, the low-energy level spacings of the SYK model become exponentially small for large $N$, rendering annealing schedules following Eq.~\eqref{eq:HQA_1} unrealistic for large-scale optimization problems with $N\gg 1$. Therefore, we now analyze annealing schedule \eqref{eq:HQA_2}, i.e., we implement time evolution under the Hamiltonian $\hat{H}_{\text{QA}}^{(2)} = s \hat{H}_{\text{MC}} + (1-s) \hat{H}_{\text{TF}}  + 4s(1-s)\hat{H}_{\text{bSYK}_q}$ with $s(t)=t/T$.

Fig.~\ref{fig:comp_4s} shows our numerical results for the same hard MaxCut graphs as in Fig~\ref{fig:comp}, with $N_{\text{bSYK}_q} = 24$ driver instances each. For $q=4$, $T^*_{\text{bSYK}_4}$ is shown in Fig.~\ref{fig:comp_4s}~(b), which features a very similar behavior as the results for schedule~\eqref{eq:HQA_1}. It therefore seems like the details of adding SYK-like fluctuations to the driver Hamiltonian play only a secondary role in the annealing outcome: In both cases~\eqref{eq:HQA_1} and~\eqref{eq:HQA_2}, significant ratios of SYK instances show enhanced performances for all considered hard graphs.

We now analyze if the performance depends on the degree of interactions $q$. In Ref.~\cite{Swingle2024}, it has been argued that for $q > 4$, the bSYK model features power-law correlation functions and an extensive low temperature entropy, akin to the fermionic SYK model. For $q=2,3$, the ground state is a spin glass; $q=4$ has been identified as a marginal case. We analyze annealing schedule~\eqref{eq:HQA_2} with $q=2$, i.e., we restrict the all-to-all interactions in the bosonic SYK Hamiltonian to two-body terms\footnote{We here only consider annealing schedule~\eqref{eq:HQA_2}. However, akin to the results for $q=4$, we expect similar results to hold for general annealing schedules.}. Results are presented in Fig.~\ref{fig:comp_4s}~(a) and (c) with orange data points. 

Successful instances of the bSYK$_2$ driver show very similar performances compared to $q=4$, see Fig.~\ref{fig:comp_4s}~(b). For MaxCut instances with moderate $ 10^4 \lesssim T^*_X\lesssim 10^6$, a seemingly larger variation of $T^*_{\text{bSYK}_2}$ leads to overall lower affected ratios, see Fig.~\ref{fig:comp_4s}~(a).
This is corroborated by the gap structure of the instantaneous eigenspectrum, which similarly features larger variations for $q=2$ compared to $q=4$ (see Appendix~\ref{sec:AppB}). 

Motivated by classical and quantum computational accessibility, next to the bSYK model there has been a recent focus on sparse versions of the (fermionic) SYK model. Here, from all original interaction terms in the Hamiltonian, only $kN$ are chosen to be present in the sparse model. Indeed, it has been shown that above a certain threshold $k_{\text{crit}} = \mathcal{O}(1)$, strong quantum chaos and holographic properties of the model survive~\cite{xu2020, Garcia2021, orman2024}. We finally study the quantum annealing performance when using sparse versions of the bSYK model, and may focus here on $q=4$ and $k=4$. Results are shown in Fig.~\ref{fig:comp_4s}~(a) and~(c). Although we consider larger annealing time windows $10 \leq T\leq 10^6$ for the sparse bSYK model, very few instances result in an enhanced annealing performance. Furthermore, $T^*$ for the sparse bSYK realizations that do show an enhancement are, in most cases, orders of magnitude larger compared to typical successful instances for the dense drivers.

\begin{figure*}
\centering
\includegraphics[width=\textwidth]{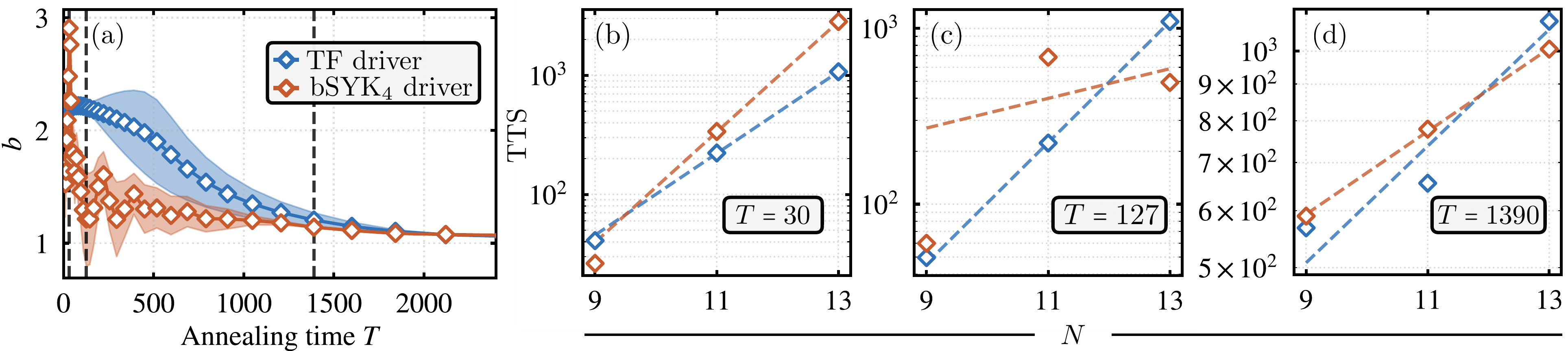}
\caption{\textbf{Solving LABS with quantum annealing.} (a) Fitted scalings of the time-to-solution TTS$\sim b^N$ for varying annealing time $T$. When fitting for a range of $N=9,11,13$, at intermediate $T$ the SYK-type driver has a scaling advantage over TF driver schedules. The shaded regions correspond to the uncertainty of the fits, which is notably large in the intermediate $T$ region. (b), (c), (d) show the TTS as a function of $N$ for three annealing times $T=30,127,1390$ [see the grey dashed lines in (a)]. Scaling fits are shown by the dashed lines.}
\label{fig:scaling}
\end{figure*}

While the dependence of the annealing performance on the degree of interactions $q$ was found to be negligible, sparsifying the bSYK Hamiltonian significantly weakens the driver's capabilities to navigate through complicated energy landscapes. We therefore speculate that is it primarily the non-locality of the SYK-type drivers that lead to more efficient quantum annealing solutions. Nevertheless, it is an interesting future research direction to study in more depth the connections between annealing performances and quantum scrambling and chaos.

From a more practical point of view and assuming that all-to-all interactions can be engineered in hardware setups, our perspective is that a cyclic annealing algorithm leads to best results, akin to what has been proposed in~\cite{Wang2022, zhang2024cyclic}: A given graph is repeatedly annealed with randomly drawn bSYK$_q$ instances, where as many terms of the bSYK model as feasible are included in the time evolution. From a large number of anneal results, the bit string that corresponds to the lowest classical energy is then chosen as the annealing solution. We found that the performance only weakly depends on how the bSYK model is included in the driver, i.e., the exact functional form of the annealing schedule is only of minor importance and can be adapted to what may be experimentally most feasible. 

\section{LABS}
\label{sec:labs}
Though finding optimal solutions of MaxCut problems is generally an $NP$-hard problem, solutions can be found very efficiently for typical instances. In contrast, the low autocorrelation binary sequence (LABS) problem does not feature such efficient heuristic algorithms. LABS was originally developed to reduce the peak power of radiation and sonar pulses~\cite{Boehmer1967, Schroeder1970, Golay1972, Packebusch_2016}, which has been used to e.g. optimize interplanetary radar measurements of spacetime curvature~\cite{Shapiro1968}. The objective of the LABS optimization is to minimize the so-called sidelobe energy of a binary sequence, which can be found by finding the ground state of the classical Hamiltonian
\begin{equation}
    \hat{H}_{\text{LABS}} = \sum_{j = 1}^{N-1} \hat{\mathcal{C}}^2(j) = \sum_{j = 1}^{N-1} \left(\sum_{i=1}^{N-j} \hat{\sigma}^z_i \hat{\sigma}^z_{i+j}\right)^2.
    \label{eq:H_labs}
\end{equation}
Here, $\hat{\mathcal{C}}(j) = \sum_{i=1}^{N-j} \hat{\sigma}^z_i \hat{\sigma}^z_{i+j}$ is the autocorrelation operator at distance $j$. The Hamiltonian Eq.~\eqref{eq:H_labs} features long-range 4-body interactions, but is fully deterministic, i.e., there is no disorder as in glassy systems. Nevertheless, the energy landscape is highly complex, and the system has similarities with disordered systems (i.e. it features self-induced disorder)~\cite{Krauth1995}. At high temperatures, variations of the replica method allow for an analytic treatment of the system~\cite{Bouchaud1994, Marinari_1994}. However, at low energy (and in particular in the ground state at $T=0$), exact properties are not known.  

All known classical algorithms feature an exponential run time in system size, and getting exact (approximate) solutions are computationally possible for only $N\leq 66$ ($N\lesssim 200$). The best known exact (heuristic) algorithm\footnote{For a full survey of classical algorithms, see in particular the supplementary materials of Ref.~\cite{Shaydulin2024}.} scales with system size as $1.73^N$ ($1.34^N$)~\cite{Packebusch_2016}. In Ref.~\cite{Shaydulin2024}, the LABS problem was studied using QAOA. Indeed, it was shown that when combining a constant depth QAOA routine with quantum minimum-finding, the scaling of the time-to-solution (TTS) with system size is $1.21^N$, which is lower than the best known classical heuristic.

Here, we study the performance of quantum annealing when applied to the LABS problem. We again focus on varying annealing times and corresponding time-to-solutions, and compare their scaling with $N$ when using transverse field and SYK-type driver Hamiltonians. To approximate the scaling of the annealing algorithm as $N$ is varied, we calculate for each annealing time $T$ the time-to-solution (see e.g.~\cite{Zhou2020}),
\begin{equation}
    \text{TTS} \propto \frac{T}{\log\big(1-p_{\text{GS}}(T)\big)}.
\end{equation}
Here, $p_{\text{GS}}$ is the probability to be in the ground state of the LABS Hamiltonian by the end of an annealing schedule of length $T$. We note that solutions to the LABS problem have varying degrees of degeneracy ($D$) for varying system size. Therefore, we define $p_{\text{GS}} = \sum_{i=1}^D p_{\text{GS}}^{(i)}$, with $p_{\text{GS}}^{(i)}$ the wave function's weight on the $i$'th degenerate bit string of the classical Hamiltonian. 

For varying annealing time $T$, we calculate the TTS for systems of sizes $N = 9,11,13$. An exponential fit to the TTS as a function of $N$ then yields the scaling $\text{TTS}\sim b^N$. The dependency $b(T)$ is shown in Fig.~\ref{fig:scaling}~(a). For three different annealing times, we show the dependence of TTS on $N$ together with the scaling fits in Fig.~\ref{fig:scaling}~(b)-(d). In particular for intermediate times, the fits are very limited in accuracy, in particular as only three data points have been used. Comparisons between the two different drivers should thus be interpreted as rough trends: While in the limit of long annealing times both TF and bSYK$_q$ drivers reach the same value of $b\sim 1.1$, at short times the scaling of the bSYK$_q$ driver is advantageous over the TF schedule\footnote{We note that we here focus on annealing schedule defined in Eq.~\eqref{eq:HQA_1} with $q=4$; however, akin to the above discussion, we expect similar results to hold for other annealing schemes.}.

While the TTS scaling of the two different drivers can be compared, we stress that a meaningful comparison to classical heuristic algorithms is impossible with our considered system sizes. To extract the scaling of the classical algorithms, $N>40$ spins were used, whereas our system sizes are significantly smaller. Larger systems and more data points are needed for the annealing results to see which algorithm decisively performs better at scale. Nevertheless, our results corroborate that the use of non-local, chaotic Hamiltonians in quantum annealing can lead to a computational speedup compared to standard TF drivers when solving classical optimization problems. 

\section{Discussion}
We have numerically studied the scenario of adding chaotic driver Hamiltonians to quantum annealing protocols, and compared resulting performances in a variety of settings to the standard transverse field.
In particular, we focused on hard instances of classical optimization problems, and showed that chaotic drivers can indeed significantly enhance the annealing performance when compared to plain transverse field annealing. 

While not all instances of SYK Hamiltonians were successful in reducing the annealing time, those that do show an advantage were seen to reach high ground state populations independent on the instantaneous minimum energy gap when using the transverse field. For practical applications of large-scale optimization problems, it is an interesting question how these success rates scale as a function of graph size $N$. Furthermore, it could be insightful to analyze the performance of other quantum alhorithms, e.g. QAOA, when using chaotic driver Hamiltonians. 

For the MaxCut problem on regular graphs, we have found comparable results for varying degree of interactions $q$, while sparsifying the Hamiltonian matrix led to significantly less efficient annealing performances. It is an interesting future research direction to study if any more direct connections between our numerical observations and the quantum chaotic and fast scrambling nature of the SYK model, as well as the degree of sparsity can be drawn. 

For practical quantum annealing setups, our results suggest that implementing additional non-local fluctuations of degree $q\geq 2$ to the transverse field lead to an enhanced performance, even if implementing the fully dense bSYK model is currently out of reach. With recent advances in the control and coherence of neutral atom quantum processors, a digitized annealing approach that includes SYK-like Hamiltonians during trotterized time evolution may constitute a promising path towards practical quantum advantage. \\

\textbf{Acknowledgements.} We thank Annabelle Bohrdt, Pietro Bonetti, Jordan Cotler, Oriana Diessel, Edward Farhi, Jernej Rudi Fin\v{z}gar, Fabian Grusdt, Alex Kamenev, Christian Kokail, Kunal Marwaha, Nishad Maskara, Rhine Samajdar, Matthias Troyer, and William Witczak-Krempa for fruitful discussions. H.S. acknowledges funding by the Deutsche Forschungsgemeinschaft (DFG, German Research Foundation) under
Germany’s Excellence Strategy—EXC-2111—390814868
and by the European Research Council (ERC) under the
European Union’s Horizon 2020 research and innovation
programme (grant agreement number 948141 — ERC
Starting Grant SimUcQuam). S.S. was supported by the U.S. National Science Foundation grant No. DMR-2245246 and by the Simons Collaboration on Ultra-Quantum Matter which is a grant from the Simons Foundation (651440, S.S.).

\section*{Appendix}
\appendix

\begin{figure*}
\centering
\includegraphics[width=1\textwidth]{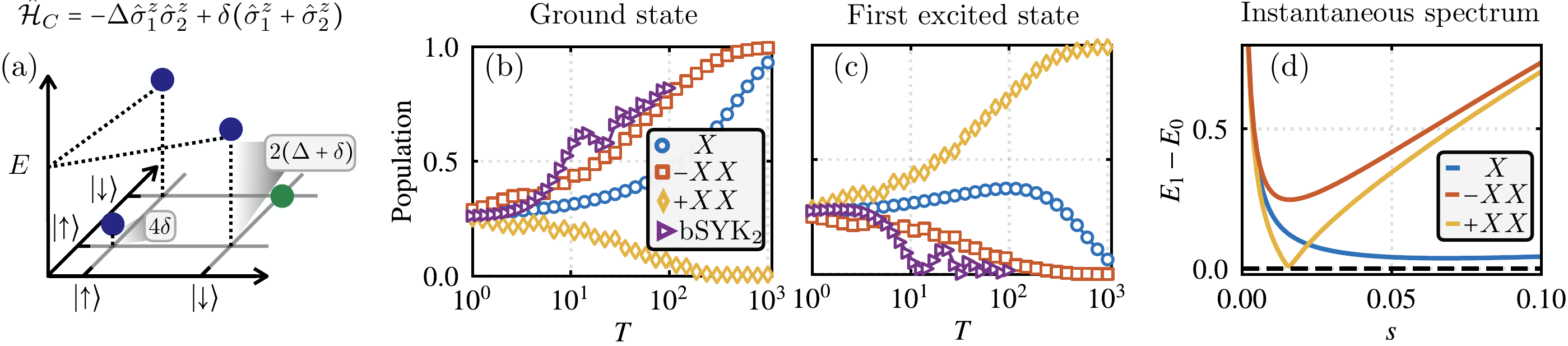}
\caption{\textbf{2-qubit toy model.} (a) Energy landscape of Hamiltonian Eq.~\eqref{eq:H_2spin_static} for $\Delta/\delta \gg 1$. Relative to the ground state $\ket{\downarrow \downarrow}$, states with Hamming distance $d_H = 1$ are largely gapped out, while a small splitting to $\ket{\uparrow \uparrow}$ exists. (b) Ground state population after annealing schedule Eq.~\eqref{eq:H_2spin} when including only the TF driver (blue circles), and when addition additional terms $\pm XX$ (yellow diamonds and orange squares, respectively). Purple triangles show results for a particular instance of the bSYK$_2$ model. (c) The same as (b) for the first excited state. (d) Instantaneous eigenspectrum showing the gap to the first excited state, $E_1 - E_0$, for the different annealing scenarios. While the $-XX$ driver gaps out the spectrum, a gap closing exists for $+XX$.}
\label{fig:toy}
\end{figure*}

\section{Toy model}
\label{sec:App1}
In the following, we construct a simple energy landscape example that illustrates the advantage of $(i)$ non-localilty and $(ii)$ additional quantum fluctuations - we explicitly thank Jordan Cotler for his suggestions. To illustrate how non-local interactions can enhance the performance of quantum annealing, consider a system of two qubits and a classical energy landscape described by the Hamiltonian
\begin{equation}
    \hat{\mathcal{H}}_{\text{C}} = -\Delta \hat{\sigma}^z_1 \hat{\sigma}^z_2 + \delta(\hat{\sigma}_1^z + \hat{\sigma}_2^z),
    \label{eq:H_2spin_static}
\end{equation}
with $\hat{\sigma}^z_i$ the Pauli-$z$ matrix on site $i$. For $\Delta/\delta \gg 1$, the energy landscape consists of sharp peaks for states $\ket{\uparrow \downarrow}$ and $\ket{\downarrow \uparrow}$, while $\ket{\downarrow \downarrow}$ is the ground state with a gap of $4\delta$ to the first excited state $\ket{\uparrow \uparrow}$ [see Fig.~\ref{fig:toy}~(a)]. The transverse field ground state, $\ket{++} = \frac{1}{2} (\ket{\uparrow \uparrow} + \ket{\downarrow \downarrow} + \ket{\uparrow \downarrow} + \ket{\downarrow \uparrow})$, has an equal weight on all four basis states. When using solely the transverse field driver (which connects states with Hamming distance $d_H = 1$), dynamically shifting the weight to the ground state requires to pass through the spiky peaks. On the other hand, when additionally allowing fluctuations of the form $\hat{\sigma}^x_1 \hat{\sigma}^x_2$, a direct connection between the two low-lying states $\ket{\uparrow \uparrow}$ and $\ket{\downarrow \downarrow}$ exists, which can help to find the ground state faster. 

We first focus on the following annealing path akin to schedule~\eqref{eq:HQA_2},
    \begin{equation}
        \hat{\mathcal{H}}_{\text{QA}} = s \hat{\mathcal{H}}_{\text{C}} + (1-s) \hat{\mathcal{H}}_{X} \pm 4s(1-s)\hat{\sigma}_1^x \hat{\sigma}_2^x,
        \label{eq:H_2spin}
    \end{equation}
where we choose additional drivers $\pm \hat{\sigma}_1^x \hat{\sigma}_2^x$. The above schedule starts in the ground state of the TF Hamiltonian; additional 2-point interactions are then added smoothly to the Hamiltonian. At $s=1/2$, the strength of both the TF and $\pm \hat{\sigma}^x_1 \hat{\sigma}^x_2$ terms are equal. At $s=1$, the Hamiltonian corresponds to the classical model  Eq.~\eqref{eq:H_2spin_static}. We solve the time-dependent Schrödinger equation governed by Eq.~\eqref{eq:H_2spin} of two spins with $\Delta = 10^3$, $\delta = 10^{-1}$, and annealing time $T$ (with parameterization $s = t/T$).

The ground state [first excited state] population after the schedule with annealing time $T$ for the different paths is shown in Fig.~\ref{fig:toy}~(a) [Fig.~\ref{fig:toy}~(b)]. 
Without additional driver Hamiltonians [i.e. a schedule given by Eq.~\eqref{eq:HQA}], the population of both $\ket{\uparrow \uparrow}$ and $\ket{\downarrow \downarrow}$ slowly rise simultaneously, before splitting at larger annealing times and reaching the adiabatic regime at times of the order of $T = \mathcal{O}(100)$ [blue dots in Figs.~\ref{fig:toy}~(a) and (b)]. This is corroborated in the instantaneous eigenspectrum as a function of annealing parameter $s$, shown in Fig.~\ref{fig:toy}~(d). A small instantaneous energy gap $\Delta_X \sim 0.04$ leads to long annealing times where the process is adiabatic, i.e. $T_X \gtrsim 600$. In contrast, when adding the additional driver  $- \hat{\sigma}^x_1 \hat{\sigma}^x_2$, the energy gap of the instantaneous eigenspectrum is significantly opened ($\Delta_{-XX} \sim 0.25$), leading to a faster entrance into the adiabatic regime for $T_{-XX} \sim \mathcal{O}(10)$ [orange squares in Fig.~\ref{fig:toy}~(b)]. We note that the annealing performance of the additional driver strongly depends on its sign: For $\hat{\mathcal{H}}_D = + \hat{\sigma}^x_1 \hat{\sigma}^x_2$, a gap closing emerges at $s\sim 0.02$, leading to a large population of the first excited state instead (while the ground state population depletes) - see the yellow diamonds in Fig.~\ref{fig:toy}~(c).

\begin{figure*}
\centering
\includegraphics[width=0.55\textwidth]{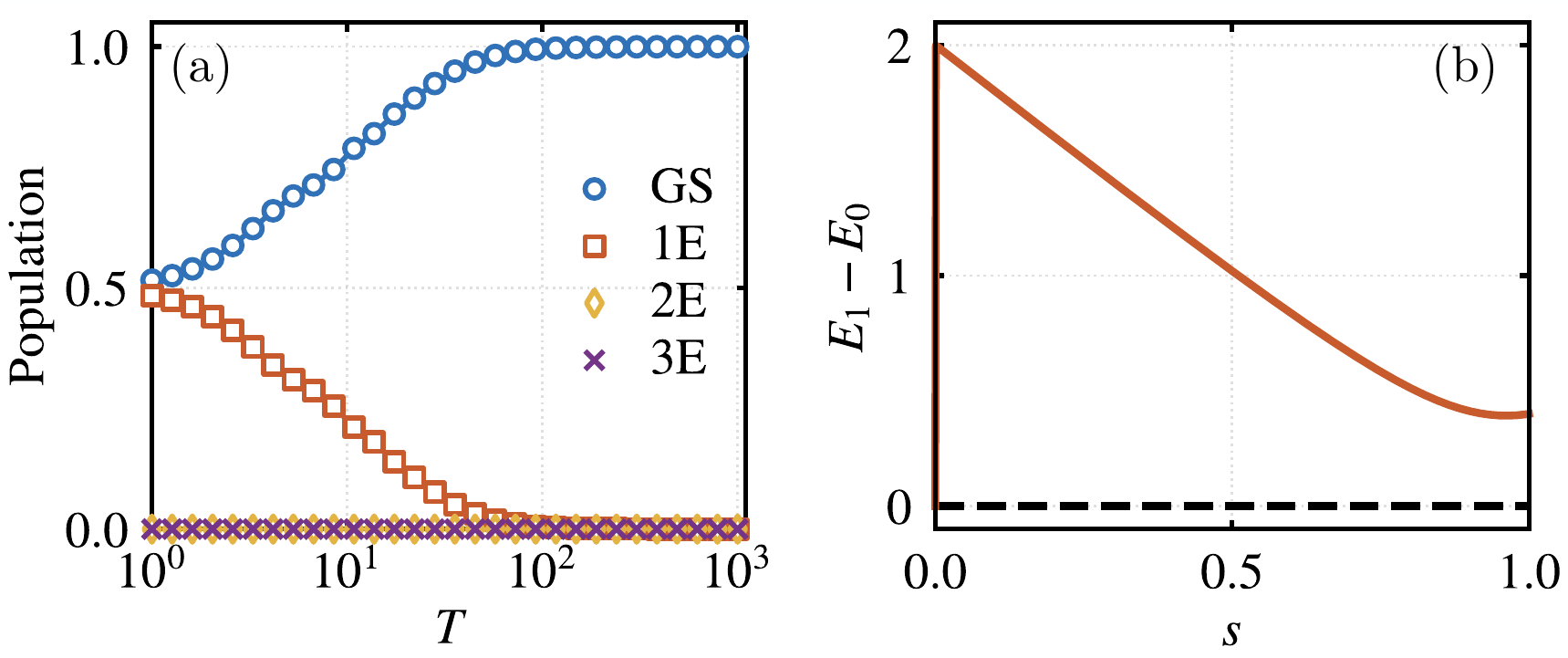}
\caption{\textbf{2-qubit toy model.} (a) Ground state and excited state populations after the annealing schedule using Eq.~\eqref{eq:H_2spin_2}. The initial state has weight only on the two low-energy states, which are directly coupled by the driver Hamiltonian. We note that in this case, using both signs ($\pm \hat{\sigma}_1^x \hat{\sigma}_2^x$) in the annealing schedule leads to identical results. (b) Instantaneous eigenspectrum. While the state at $s=0$ is two-fold degenerate, the spectrum features a large energy gap for any $0<s\leq 1$.}
\label{fig:toy2}
\end{figure*}

The sign of the interactions are seen to play an important role in the above example. In a more general setting, adding additional terms to the driver with randomly assigned weights may lead to further quantum interferences that result in a faster convergence to the ground state. For the simple 2-qubit landscape, we replace the $\hat{\sigma}^x_1\hat{\sigma}^x_2$ term by
\begin{equation}
    \sum_{\alpha_1, \alpha_2} J^{\alpha_1, \alpha_2} \hat{\sigma}^{\alpha_1}_1\hat{\sigma}^{\alpha_2}_2,
\end{equation}
which precisely corresponds to the bosonic SYK Hamiltonian Eq.~\eqref{eq:bSYK} for $q=2$ and $N=2$ spins. Indeed, for a subset of randomly drawn prefactors, the annealing performance can be further enhanced, as illustrated in Fig.~\ref{fig:toy} (purple data). The (bosonic) SYK model with its random combination of $q$-body spin interaction terms thus emerges as a natural candidate to study in quantum annealing protocols.

Lastly, we look at performances when annealing schedule~\eqref{eq:HQA_1},
    \begin{equation}
\hat{\mathcal{H}}_{\text{QA}} = s \hat{\mathcal{H}}_{\text{C}} \pm (1-s)\hat{\sigma}_1^x \hat{\sigma}_2^x.
        \label{eq:H_2spin_2}
    \end{equation}
Here, we start in the ground state of the Hamiltonian $\pm \hat{\sigma}^x_1 \hat{\sigma}^x_2$ and linearly anneal to the classical Hamiltonian. 
We note that the ground state of $\pm \hat{\sigma}_1^x \hat{\sigma}_2^x$ is two-fold degenerate. In the following, we choose as initial states $\ket{\uparrow \uparrow} \pm \ket{\downarrow \downarrow}$. Annealing performances for varying $T$ are shown in Fig.~\ref{fig:toy2}~(a).

\begin{figure*}
\centering
\includegraphics[width=0.65\textwidth]{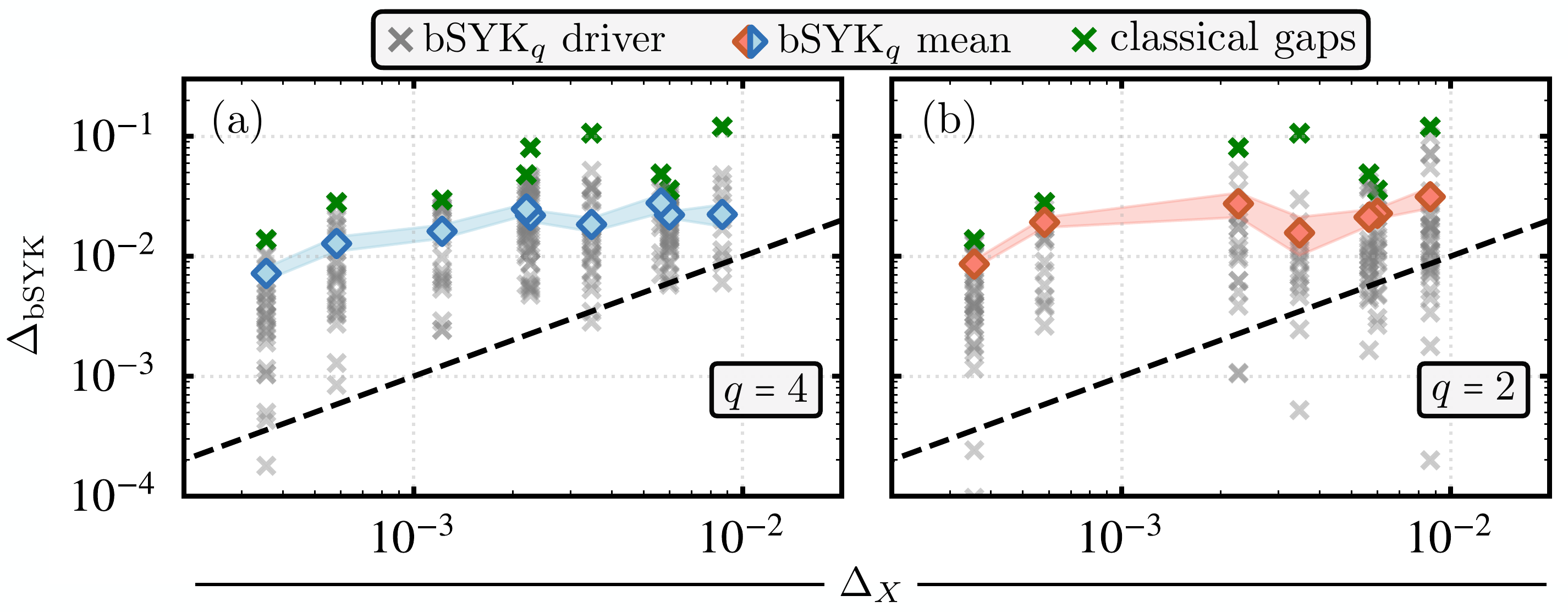}
\caption{\textbf{Minimum instantaneous energy gaps.} Comparison of minimum instantaneous energy gaps when using the TF and bSYK$_4$ driver~\eqref{eq:HQA_1} for $q=4$ (a) and $q=2$ (b). (a) is the same as in Fig.~\ref{fig:comp_gap}, and is reprinted here for easier comparison. Grey crosses show individual instances of the SYK-type driver, solid diamonds show corresponding mean values (with their error given by the shaded lines). Green crosses indicate the classical gaps at $s=1$, which is an upper bound for the instantaneous gap $\Delta_{\text{bSYK}_q}$. Due to different conserved symmetries, $\Delta_{\text{bSYK}_q} = E_2 - E_{\text{GS}}$, while $\Delta_{X} = E_1 - E_{\text{GS}}$ (see main text). While gaps are significantly opened in both cases, $q=2$ features notably wider distributions in particular for problems with moderately small gap sizes $\Delta_X$.}
\label{fig:gaps_q}
\end{figure*}

Due to the structure of the initial state, the ground state population at $t=0$ has a large overlap with the target state of $50\%$. The driver Hamiltonian then directly couples the two populated states, leading to a fast annealing to the ground state. This is corroborated in Fig.~\ref{fig:toy2}~(b), which shows that the spectrum (apart from the degeneracy at $t=0$) is fully gapped out throughout the annealing schedule. We note that in the above example, a combination of large overlap of the initial state with the target state together with the introduced coupling constitutes a highly idealized scenario for quantum annealing. Nevertheless, it underlines and motivates our perspective that non-local terms can help to navigate through spiky energy landscapes.

\section{Minimum instantaneous gaps}
\label{sec:AppB}

In the main text, we have compared the minimum energy gaps of the instantaneous eigenspectrum of the bSYK annealing path~\eqref{eq:HQA_1} to the TF driver~\eqref{eq:HQAX} for $q=4$. Fig.~\ref{fig:gaps_q} shows a comparison between $q=4$ (a) and $q=2$ (b). As in the main text, we focus on $\Delta_{\text{bSYK}_q} = \text{min}_s (E_2 -E_{\text{GS}})$ and $\Delta_{X} = \text{min}_s (E_1 -E_{\text{GS}})$ due to the different conserved symmetries. While $\Delta_{\text{bSYK}_q}>\Delta_{X}$ for most instances with both $q=2$ and $q=4$, the case of two-body interactions has notably larger variations in particular for moderately small gaps $\Delta_X$. This is in line with results of $T^*$ [see Fig.~\ref{fig:comp_4s}], where larger variations for $q=2$ lead to smaller success ratios within the considered annealing time windows.

\bibliography{annealing}

\end{document}